\documentclass[amsmath,amssymb,twocolumn,showpacs,showkeys,superscriptaddress]{revtex4}

\usepackage{graphicx}

\usepackage{bm}

\usepackage{amssymb}

\begin{document}

\title{Influence of geography on language competition}

\author{Marco Patriarca}
  \email{marco.patriarca[at]gmail.com}

\author{Els Heinsalu}
  \email{els[at]kbfi.ee}

\affiliation{National Institute of Chemical Physics and Biophysics, R\"avala 10, 15042 Tallinn, Estonia}
\affiliation{Institute of Physics, University of Tartu, T\"ahe 4, Tartu 51010, Estonia}

\date{\today}

\begin{abstract}
  Competition between languages or cultural traits diffusing in the same geographical area is studied
  combining the language competition model of Abrams and Strogatz and a human dispersal model on an inhomogeneous substrate.
  Also, the effect of population growth is discussed.
  It is shown through numerical experiments that the final configuration of the surviving language can be strongly affected by geographical and historical factors.
  These factors are not related to the dynamics of culture transmission, but rather to initial population distributions as well as geographical boundaries and inhomogeneities, which modulate the diffusion process.
\end{abstract}



\maketitle


\section{Introduction}
Currently, statistical mechanics and stochastic models are employed to study a wide range of topics,
not only in physical sciences,
but also in interdisciplinary applications in biology as well as in social and historical sciences.
Examples include applications to financial time series~\cite{Bouchaud2005a},
population dynamics~\cite{Murray2002a}, and archeology~\cite{Ammerman1984a}.
Recently, also various problems in linguistics have been approached
using methods imported from the theory of complex systems and statistical mechanics;
see Refs.~\cite{Stauffer2005a,Schulze2006c,Wichmann2008b} for an overview.

In the evolution and dispersal of biological species the importance of geography is well known~\cite{Lomolino2006a}.
The goal of the present paper is to study how purely geographical and historical constraints
can affect the dynamics of different languages or cultural traits competing in a region.
Previously the influence of geography on language dynamics has been investigated e.g. in Refs.~\cite{Patriarca2004c,Schulze2007a,Stauffer2007a,Hadzibeganovic2008a}.

We start from the Abrams-Strogatz (AS) model~\cite{Abrams2003a} of two fixed competing languages.
The term ``fixed'' refers to the fact that the evolution of language is neglected on the time scale considered,
so that the model is formally similar to a model of population dynamics of two biological species.
By ``competing'' one means that at any time speakers can switch to the other language,
as a consequence of the interaction between speakers of language 1 and 2.
In order to take into account population growth, dispersal, and the effect of geographical inhomogeneities,
we introduce in Sec.~\ref{model} a more general model.
This model is then applied to some (idealized) examples
concerning the influence of initial conditions (Sec.~\ref{ic}), boundary conditions (Sec.~\ref{bc}),
and geographical barriers (Secs.~\ref{barrier} and \ref{island}).
These examples show how extending a 0-dimensional (i.e. homogeneous) model of a culture transmission 
to physical space gives rise to new, unexpected effects.
Results are summarized in Sec.~\ref{conclusion}.

\section{The model}
\label{model}

In this section we formulate a  model of culture dispersal and competition.
We are interested in a description regarding a historical context\footnote{Our goal in this paper is not to reconstruct any real historical situation.},
in which two populations using different languages (or with different cultural traits) 1 and 2 disperse across a region.
Simultaneously, the competition between the languages takes place and possibly the populations also grow.
As a starting model we assume  the AS language competition model.
A diffusion term, describing population dispersal, and an advection term, taking into account geographical inhomogeneities, are then added; population growth is taken into account through a logistic term.

\subsection{The Abrams-Strogatz model}
\label{sec-as}

The AS model was first introduced in Ref.~\cite{Abrams2003a} 
in order to describe the time evolution of the population size of various endangered languages.
The model can be formulated through the following equations,
\begin{eqnarray} \label{AS}
  \frac{dN_1}{dt} &=&
  R(N_1,N_2) =
  \frac{s_1}{\tau} \, N_1^a \, N_2
  -    \frac{s_2}{\tau} \, N_2^a \, N_1 \ ,
  \nonumber \\
  \frac{dN_2}{dt} &=&
  - R(N_1,N_2) =
  -  \frac{s_1}{\tau} \, N_1^a \, N_2
  +    \frac{s_2}{\tau} \, N_2^a \, N_1 \ .
\end{eqnarray}
Here $N_i(t)$ ($i=1,2$) represents the fraction of speakers of language $i$.
The quantity $k_1 = s_1/\tau$ is the rate constant for the switch of a speaker of language 2 to language 1, and vice versa for $k_2 = s_2/\tau$.
The dimensionless parameter $s_i$, referred to as \emph{language status},
represents the \emph{attractiveness of language} $i$; 
it can result from a combination of factors, such as, e.g., the language prestige and usefulness.
Here we follow the normalization convention $s_1 \!+\! s_2 \!=\! 1$.
As a consequence $s_1/\tau \!+\! s_2/\tau \!\equiv\! 1/\tau$,
i.e., $\tau$ represents the time scale.
The coefficient $a$ varies slightly around the value $a \!=\! 1.3$ for different languages, 
as found in Ref.~\cite{Abrams2003a}; in this article the value $a = 1.3$ is assumed.

In the analogy with a population dynamics model it should be noticed that the reaction term $R(N_1,N_2)$ in Eqs.~(\ref{AS})
contains a positive and a negative contribution, depending on both populations $N_1$ and $N_2$, which represent an advantage and disadvantage due to the encounter with an individual of the other ``species''.
In other words, speakers of population 1 and 2 behave symmetrically to each other as prey and predator at the same time~\cite{Murray2002a}.
While this situation is not usual in biology, 
it can be justified for the interaction between two cultural traits~\cite{Abrams2003a}.

The analysis shows that the AS model has one unstable and two stable equilibrium points.
The latter ones correspond to one of the languages surviving and the other one disappearing.
Which final state will be reached depends on the initial populations $N_i(t_0)$, 
the status parameters $s_i$, 
as well as on the value of $a$.
The critical values of parameters defining the unstable equilibrium point can be obtained 
from Eqs.~(\ref{AS}) setting the rate term $R$ equal to zero,
\begin{equation} \label{cond}
  \frac{N_1^*}{N_2^*} = \left(\frac{s_2}{s_2}\right)^{1/(a-1)} \, .
\end{equation}
When the ratio $N_1^*/N_2^*$ is larger than the right hand side of condition~(\ref{cond}) at some time $t \!=\! t'$, 
then $R(N_1(t),N_2(t)) \!>\! 0$ at any later time $t > t'$ and $N_2 \!\to\! 0$ for $t \!\to\! \infty$, 
while $N_1\!\to\! N' \!\equiv\! N_1(t') \!+\! N_2(t')$.
The opposite takes place if $N_1^*/N_2^*$ is smaller than the right hand side.

\begin{figure*}[ht]
 \centering
 \includegraphics[width=13.0cm]{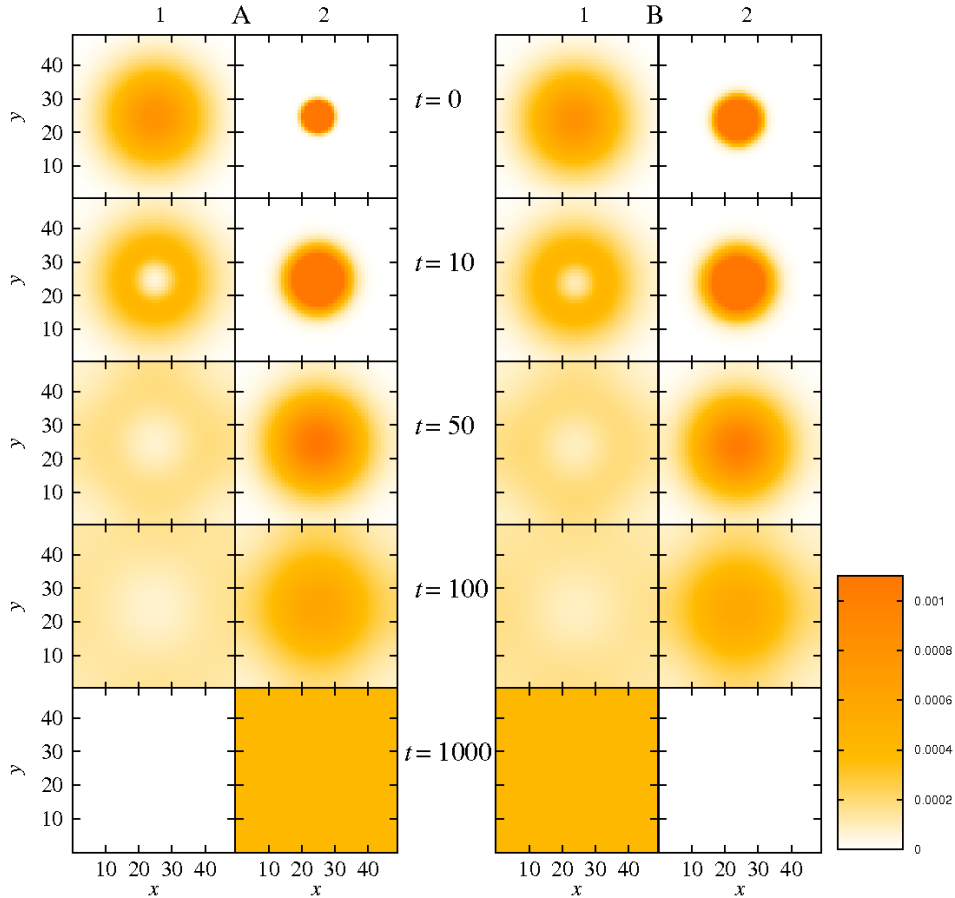}
 \caption{
 \label{multiplot-IC1}
 Comparison of the evolution of population densities $f_1(x,y,t)$ and $f_2(x,y,t)$ (columns 1 and 2) 
 for two languages with status $s_1 \!=\! 1 \!-\! s_2 \!=\! 0.55$ 
 for different widths of the initial distribution $f_2(x,y,0)$.
 The initial distributions $f_i(x,y,0)$ are  given by Eq.~(\ref{fi}).
 Example A: a more localized initial distribution $f_2(x,y,0)$ with $\sigma_1 \!=\! 1.75$. 
 Example B: a more spread $f_2(x,y,0)$ with $\sigma_1 \!=\! 3$.
 All other parameters are the same, see text for details.
}
\end{figure*}

\subsection{Generalized model}
Population and culture spreading may be affected by a wide range of geographical factors,
due to physical barriers such as water boundaries and mountains or
e.g. geophysical features such as type of ground and spatial distribution of resources~\cite{Lomolino2006a}.
While the underlying mechanisms determining the influence of such geographical factors are in general complex,
in a first approximation their overall effect can be described statistically.
In fact, dispersal of human populations in an geographical environment
recalls the diffusion of Brownian particles modulated by an external field or an inhomogeneous substrate.
For example, human dispersal in neolithic Europe~\cite{Davison2006a} and 
during the early colonization of South-America~\cite{Martino2007a} has been studied
using advection-diffusion equations with a logistic term taking into account population growth
(i.e., employing the so-called Fisher equation). 

In order to extend the AS model to take into account the geographical inhomogeneities,
we merge it with the two-dimensional geographical model of human dispersal and growth
proposed in Refs.~\cite{Davison2006a,Martino2007a}.
The corresponding evolution equations read,
\begin{eqnarray}  \label{genmodel1}
   \frac{\partial f_1}{ \partial t }
   &=& R(f_1,f_2)
   - \nabla \cdot \left( \, \mathbf{F} f_1 \right)
   +  \nabla ( D \nabla f_1 )
   \nonumber \\
   &+& \alpha f_1 \left( 1 - \frac{f_1+f_2}{K} \right) \ ,
   \nonumber \\
   \frac{\partial f_2}{ \partial t }
   &=& - R(f_1,f_2)
   - \nabla \cdot \left( \, \mathbf{F} f_2 \right)
   + \nabla ( D \nabla f_2 )
   \nonumber \\
   &+& \alpha f_2 \left( 1 - \frac{f_1+f_2}{K} \right) \ ,
\end{eqnarray}
with the reaction term given by
\begin{eqnarray}  \label{R}
R(f_1,f_2) = k \left( s_1 f_1^{\, a}  f_2 - s_2 f_2^{\, a}  f_1 \right) \, .
\end{eqnarray}
The quantity $f_i = f_i(x,y,t)$ represents the population density of speakers of language $i$,
while the constant $k$ in Eq.~(\ref{R}) is an effective rate constant and $s_i$ still represents the status of language $i$.
In Eqs.~(\ref{genmodel1}) population movement is described by the advection term containing 
the external force field $\mathbf{F}(x,y) = (F_x(x,y), F_y(x,y))$ and by the diffusion term with the diffusion coefficients $D=D(x,y,)$.
In general $\mathbf{F}$ and $D$ are related; 
their explicit form depends on the problem considered.
One possibility is to set $\mathbf{F}(x,y) = 0$ and 
describe the inhomogeneous character of the substrate 
through a space-dependent diffusion coefficient $D(x,y)$~\cite{Davison2006a,Martino2007a}.
In the model systems studied below we ascribe the inhomogeneous character of dispersal to the external force, 
$\mathbf{F} \!=\! \mathbf{F}(x,y)$, while $D$ is kept constant.
For illustrative purposes and in analogy with Brownian motion, 
the force field is expressed as the gradient of a potential, 
$\mathbf{F}(x,y) \!=\!-\!\nabla U(x,y)$.
The logistic terms with Malthus rate $\alpha$ and carrying capacity $K$ in Eqs.~(\ref{genmodel1}) 
take into account the population growth.

According to Eqs.~(\ref{genmodel1}), populations 1 and 2 disperse independently,
whereas the densities $f_1$ and $f_2$ are coupled only through 
the reaction term $R$ and through
the logistic terms, which introduce a negative competitive coupling proportional to $- f_1 f_2$.
Also, it should be noticed that Eqs.~(\ref{genmodel1}) describe two populations with identical dispersal and growth properties, 
differing only in the way how the respective language interact, according to the term $R$ given by Eq.~(\ref{R}).
As a consequence, the total population density $f = f_1 + f_2$ follows a diffusion-advection-growth process without culture transmission, obtained by summing Eqs.~(\ref{genmodel1}),
\begin{eqnarray}  
\frac{\partial f}{\partial t}
=
- \nabla \cdot [ \, \mathbf{F} f ]
+  \nabla (D \nabla f)
+ \alpha f \left( 1 - \frac{f}{K} \right) \, .
\end{eqnarray}

In order to solve Eqs.~(\ref{genmodel1}) numerically, one can approximate the derivatives through the corresponding finite differences, 
replacing the problem in the continuous time and space variables $(t, x, y)$ with the one on a lattice $(k,m,n)$
defined by the discrete variables $t_k = k \, \delta t$, $x_m = m \, \delta x$, and $y_n = n \, \delta y$, respectively,
where $k$, $m$, $n$ are integers, while $\delta t, \delta x, \delta y$ are the lattice steps.
By using the explicit Euler integration scheme~\cite{Press1986a}, 
for constant $D$ one obtains from (\ref{genmodel1}) the finite-difference equations
\begin{eqnarray}  \label{genmodel2}
   &&\frac{\delta_k f_i}{\delta t} = 
   \pm \, R(f_1,f_2)
   - \frac{\delta_m (F_x f_i)}{\delta x} - \frac{\delta_n (F_y f_i)}{\delta y}
   \nonumber \\
   &&+ D \left[ \frac{\delta_m^{\,2} f_i}{\delta x^2} + \frac{\delta_n^{\,2} f_i}{\delta y^2} \right]
   + \alpha f_i \left(1 - \frac{f}{K} \right)
   \, , ~i = 1,2.
\end{eqnarray}
Here the term $+R$ corresponds to $i=1$ and $-R$ to $i=2$.
The time difference operator $\delta_k$, when applied to a generic function $\phi(k,m,n)$, gives
$$
\delta_k \phi(k,m,n) = \phi(k+1,m,n) - \phi(k,m,n) \, .
$$ 
As for the $x$-difference operators, $\delta_m$ is defined as 
$$
\delta_m \, \phi(k,m,n) = [\phi(k,m\!+\!1,n) - \phi(k,m\!-\!1,n)]/2 \, ,
$$
while the second difference operator $\delta_m^{\, 2}$ as
$$
\delta^{\,2}_m \phi(k,m,n) = \phi(k,m\!+\!1,n) + \phi(k,m\!-\!1,n) - 2\phi(k,m,n) \, ;
$$
analogous definitions apply to the $y$-difference operators $\delta_n$ and $\delta_n^{\,2}$.
If the finite-difference equations (\ref{genmodel2}) are used to solve numerically the continuous problem,
suitable constraints have to be imposed on the values of the steps  $\delta t, \delta x, \delta y$ in order
to limit the numerical integration error~\cite{Press1986a}. 

One can also consider Eqs.~(\ref{genmodel2}) 
as a general lattice model of dispersal, growth, and cultural interaction of two populations, 
with no reference to the continuous equations (\ref{genmodel1}).
Such a discrete model is employed in the examples presented in the forthcoming sections,
even if for convenience we will refer also to the corresponding continuous limit represented by Eqs.~(\ref{genmodel1}).

\begin{figure}[ht]
 \centering
 \includegraphics[width=8cm]{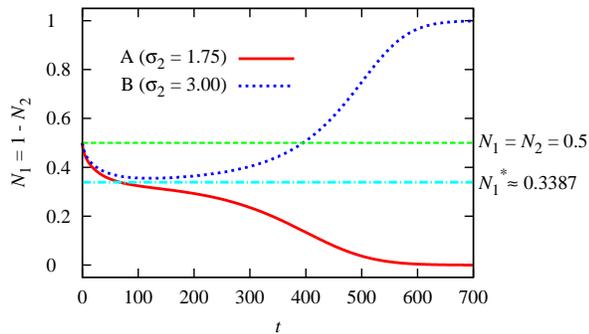}
 \caption{
  \label{average-IC1}
 Time evolution of the total population $N_1(t)=1-N_2(t)$ 
 with a status $s_1 = 0.55 = 1 - s_2$ 
 for the examples A (continuous line) and B (dotted line) of Fig.~\ref{multiplot-IC1}.
 The critical fraction $N_1^* = 0.3387$ given by the AS model for the survival of language 1 
 and the line corresponding to $N_1 = N_2$ are also drawn.
}
 \end{figure}

\section{Influence of initial conditions}
\label{ic}
%
The minimal spatial version of the AS model is obtained when taking in Eqs.~(\ref{genmodel1}) 
the  diffusion coefficient $D$ equal to a constant, and neglecting the growth and advection terms.
Similarly to the original AS model, it admits two stable equilibrium solutions, 
corresponding to one of the two languages surviving and the other one disappearing,
as can be shown by linear stability analysis.
Depending on the initial conditions, there exist also unstable equilibrium solutions.
In the presence of a local spatial noise the unstable solutions 
are washed out by the random fluctuations~\cite{Stauffer2007a}.
However, the minimal spatial model (with no noise) presents also some new effects 
respect to the corresponding homogeneous version.
In this section we discuss some examples related to the influence of initial conditions.

\subsection{Cultural interaction and dispersal without growth}

Differently from a homogeneous model, such as the AS model, 
described by ordinary differential equations, 
in the spatial model the evolution in time and space depends on 
the form of the initial population densities $f_i(\mathbf{r},t_0)$ (c.f. Sec.~\ref{sec-as}).
While this is a standard mathematical property, 
it has a relevant meaning in terms of geographical and historical conditions.
We consider the discretized Eqs.~(\ref{genmodel2}) on a square lattice with reflecting boundary conditions,
assuming $\mathbf{F}=0$ (no advection) and $\alpha=0$ (no growth).
Here and in the other simulations units are chosen to have a constant diffusion coefficient $D=1$.
Such a model represents a simplified version of a region that is isolated and geographically homogeneous.
We use a lattice of size $50 \times 50$ with $\Delta x \!=\! \Delta y \!=\! 1$; 
the time step is $\Delta t \!=\! 0.01$ and the reaction constant $k\!=\!2000$.

\begin{figure*}[ht]
 \centering
 \includegraphics[width=13.0cm]{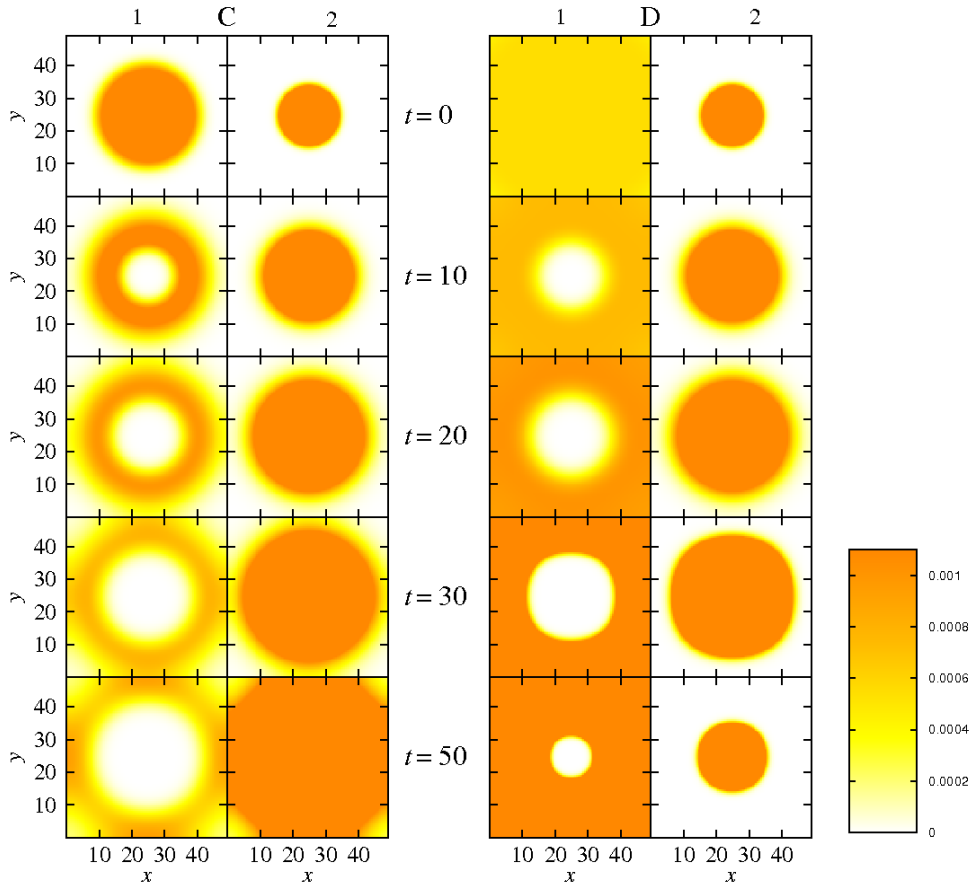}
 \caption{
 \label{multiplot-IC-growth1}
 Comparison of the evolution of population densities $f_1(x,y,t)$ and $f_2(x,y,t)$ (columns 1 and 2) 
 for two languages with status $s_1 \!=\! 1 \!-\! s_2 \!=\! 0.55$ 
 for different widths of the initial distribution $f_1(x,y,t_0)$.
 All other parameters are the same, see text.
 Example C: initial distribution $f_1(x,y,t_0)$ localized within a radius $R_1 \!=\! 15$.
 Example D: uniform initial distribution $f_1(x,y,t_0) \!=\! \mathrm{const}$.
 }
\end{figure*}

In order to check the consistency of the minimal spatial model, 
we have verified the AS model predictions by choosing uniform initial conditions, $f_i(x,y,t_0) \!= \mathrm{const}$.
In this case the diffusion terms in Eqs.~(\ref{genmodel1}) or (\ref{genmodel2}) are zero 
and the distributions remain uniform at any time $t$;
this is the only case in which integrating over the space coordinates exactly gives back the AS model (\ref{AS}).
Uniform initial conditions for $f_1$ and $f_2$ represent a historical moment when 
the populations 1 and 2 were broadly distributed across the territory.
Population 1 is observed to disappear whenever the initial density ratio $f_1/f_2$ 
is lower than the critical value $N_1^*/N_2^*$ given by condition (\ref{cond}).

\begin{figure}[ht]
 \centering
 \includegraphics[width=6.5cm]{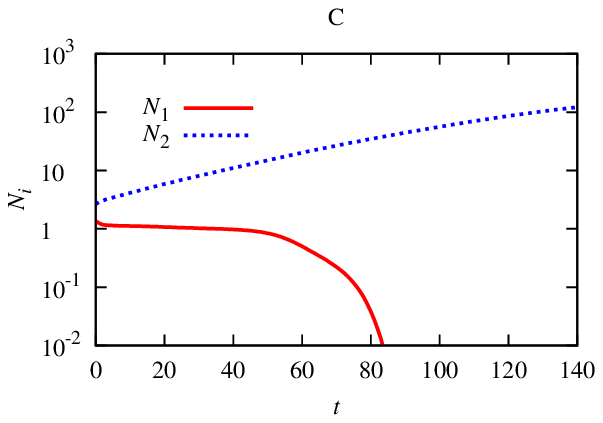}
 \includegraphics[width=6.5cm]{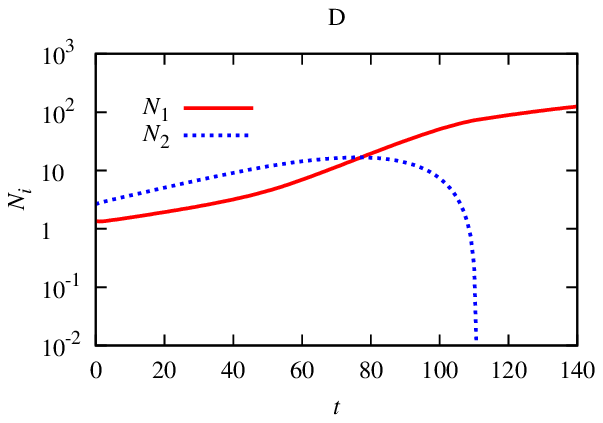}
 \caption{
  \label{average-IC-growth1}
 Time evolution of the total populations $N_1(t)$ and $N_2(t)$ 
 corresponding to the examples C (left) and D (right) of Fig.~\ref{multiplot-IC-growth1}.
}
 \end{figure}

In the non-uniform case, 
we have found for various values of parameters 
that a broader initial distribution represents a disadvantage 
when population growth is negligible ($\alpha \approx 0$).
We illustrate this effect for two languages with status $s_1 \!=\! 0.55$
and $s_2 \!=\! 1 \!-\! s_1 \!=\! 0.45$ ($s_1 > s_2$) 
and equal initial total population sizes $N_{1}(t_0) \!=\! N_{2}(t_0) \!=\! 1/2$.
Since $\alpha \!=\! 0$, the total number of speakers is conserved and for simplicity we normalize
it to one, $N_{1}(t) + N_{2}(t) \!=\! 1$. 
Thus $N_i(t)$ represents here the fraction of speakers of the $i$-th language at time $t$.
With such parameters and initial conditions,
language 1 would be clearly favored in the uniform case (or in the AS model) 
while language 2 would disappear.
However, this does not necessarily happen when space dimensions are taken into account.

Let us investigate the situation when populations 1 and 2 are initially distributed according to Gaussian densities,
\begin{equation}\label{fi}
  f_i(x,y,t_0)
  = \frac{ N_{i}(t_0) } { 2\pi\sigma_{i}^2 }
  \exp \left[ - \frac{ (x-x_i)^2 + (y-y_i)^2 } { 2\sigma_{i}^2 } \right] \, ,
\end{equation}
with $i = 1,2$, $x_i \!=\! y_i \!=\! 25$, i.e., the average positions are located in the center of the simulation area
(see Fig.~\ref{multiplot-IC1} top).
We remark that, given the symmetry of the initial configuration,
using reflecting or periodic boundary conditions leads to perfectly equivalent results. 
For population 1 we assume $\sigma_{1} \!=\! 10$ in both examples A and B, 
whereas for population 2 we assign $\sigma_{2} \!=\! 1.75$ in example A 
and $\sigma_{2} \!=\! 3$ in example B.
The particular initial configurations assumed can be interpreted from a historical point of view
as the sudden appearance of a high population density of speakers 2 
in the center (of mass) of population 1\footnote{A historical conquest scenario has been recently studied 
in the framework of the bit-string model of language evolution by Schulze and Stauffer in Ref.~\cite{Schulze2007b}.}.
One is then interested in predicting the final configuration, i.e., which language will eventually prevail.
Surprisingly, even when the two initial population sizes are equal, it is not the language with a higher status which necessarily survives.
This situation is illustrated by example A in Fig.~\ref{multiplot-IC1}.
In this case, the initial population distribution, with $\sigma_{2} \!=\! 1.75$, 
is sufficiently narrow and the associated high population density favors the $1 \!\to\! 2$ reaction,
see the function $R(f_1,f_2)$ given by Eq.~(\ref{R}).
Eventually language 1 disappears despite its higher status and the same initial population. 
Starting from a wider initial density of population 2, $\sigma_{2} = 3$,
while all other parameters maintain the same values as in example A, 
the opposite final configuration is recovered,
i.e., it is population 2 which now disappears (see example B in Fig.~\ref{multiplot-IC1}).

In Fig.~\ref{average-IC1} we compare the total populations $N_1(t) \!\equiv\! 1 \!-\! N_2(t)$ 
for the same examples A and B of Fig.~\ref{multiplot-IC1}.
As one can notice, the total population $N_1$ of example B becomes smaller than $N_2$
immediately at $t > t_0 = 0$ and remains smaller until $t \!\approx\! 400$; 
however, eventually population 1 prevails.
This is possible since language 1 has a higher status, $s_1 > s_2$.
The snapshot at time $t \!\approx\! 100$ in Fig.~\ref{multiplot-IC1} shows that
population densities have become almost uniform by that time, so that one can use the AS model
to estimate the critical ratio for the survival of language 1.
From Eq.~(\ref{cond}), one obtains $N_1^*/N_2^* \!\approx\!  0.5122$ for $s_1 = 1 - s_2 = 0.55$,
corresponding to a critical fraction $N_1^* \!\approx\! 0.3387$.
In Fig.~\ref{average-IC1}, in the example B, the surviving population $N_1(t) > N_1^*$ at any time $t$ ,
while in the example A $N_1(t)$ crosses the line $N_1 = N_1^*$, beginning its irreversible decrease.

\begin{figure*}[ht]
 \centering
 \includegraphics[width=13.0cm]{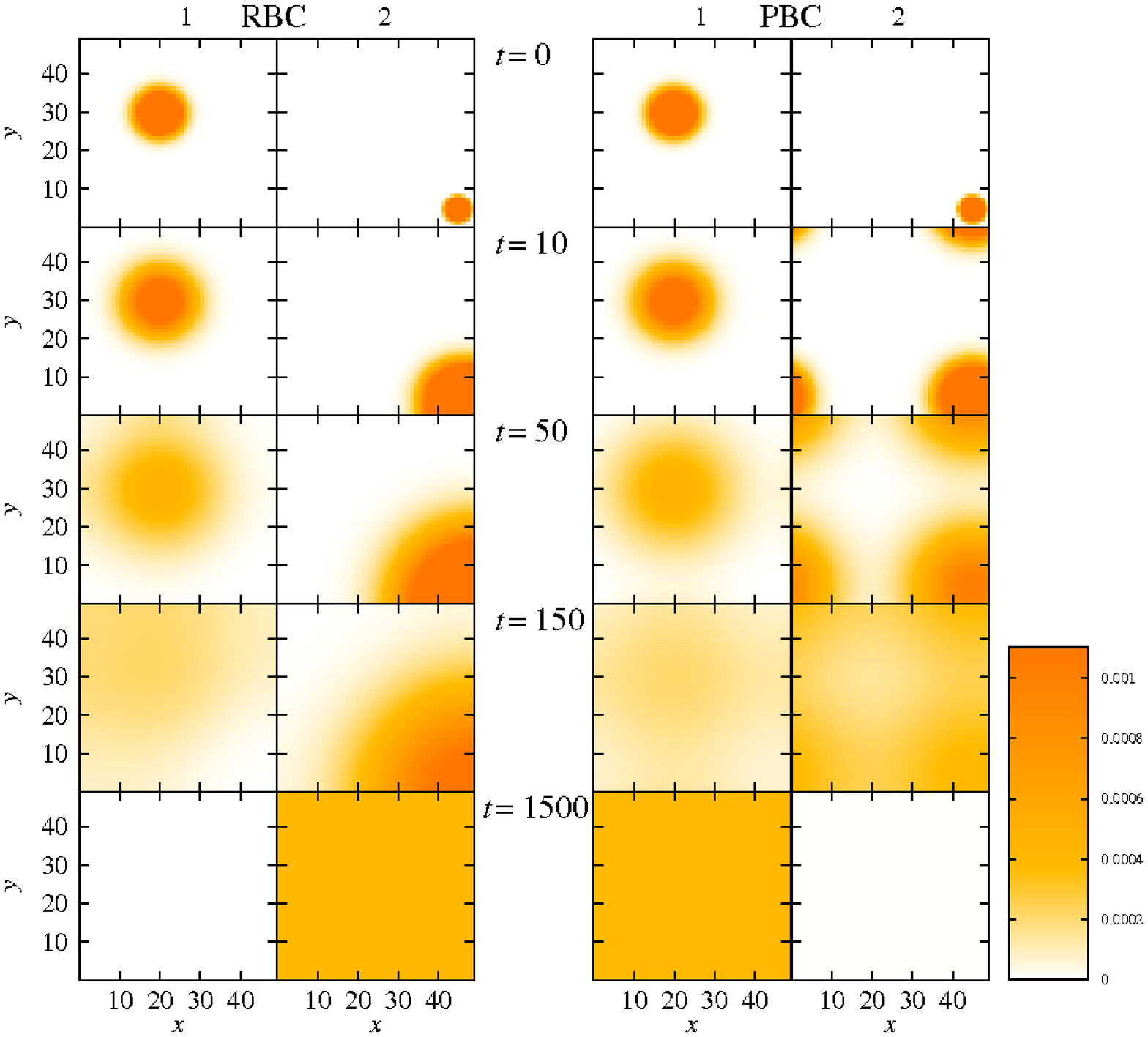}
 \caption{
 Comparison of the evolution of population densities $f_1(x,y,t)$ and $f_2(x,y,t)$ (columns 1 and 2 respectively) with language status $s_1 \!=\! 0.45 \!=\! 1 \!-\! s_2$ for reflecting (RBC, left part) and periodic (PBC, right part) boundary conditions, while all other parameters are the same.
 See text for details.
 }
 \label{bc-multiplot}
\end{figure*}

\subsection{Cultural interaction and dispersal with growth}
In this subsection we show that taking into account population growth, 
for sufficiently large values of $\alpha$,
a larger initial spreading can lead to the survival of the language,
on the contrary to the situation with no growth, considered above.
In Fig.~\ref{multiplot-IC-growth1} we present two illustrative examples, C and D,  
which differ from each other with respect to the initial spread of population 1.

\begin{figure}[ht]
 \centering
 \includegraphics[width=8.0cm]{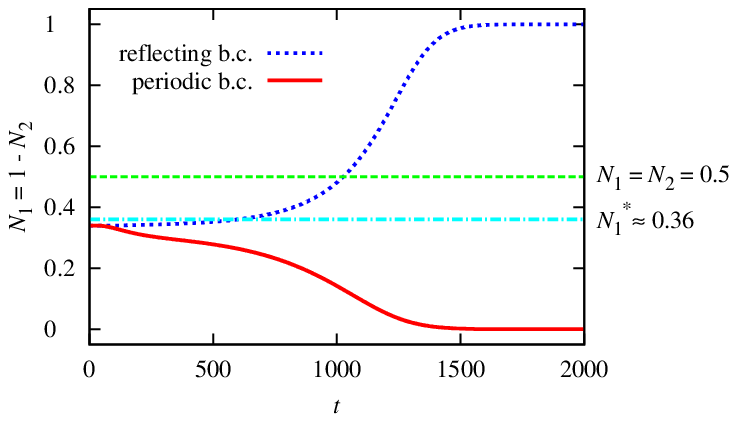}
 \caption{\label{bc-average}
 Time evolution of the total population $N_1(t) \!=\! 1 \!-\! N_2(t)$ 
 for the examples RBC and PBC of Fig.~\ref{bc-multiplot}.
 }
\end{figure}

Performing the numerical simulations, we have used a $100 \!\times\! 100$ simulation area 
with $\Delta x \!=\! \Delta y \!=\! 0.5$ and periodic boundary conditions.
The time step is $\Delta t \!=\! 0.001$ and the reaction constant $k \!=\! 1000$.
Regarding the growth term, a rate $\alpha \!=\! 0.03$ and a carrying capacity $K \!=\! 0.1$ have been used.
The initial densities have been chosen of the form
\begin{equation}\label{fi2}
  f_i(x,y,t_0) =
  \left\{
  \begin{array}{ll}
    \mathcal{N}_i \, , &r_i(x,y) < R_i \, , \\
    \frac{2 \, \mathcal{N}_i} 
    { 1 \, + \, \exp\left\{ - [ \, r_i(x,y) - R_i \, ]^2 / \, 2 \sigma_i^2 \, \right\} }
    \, , &r_i(x,y) > R_i \, ,
  \end{array}
  \right.
\end{equation}
where $r_i(x,y) \!=\! \sqrt{(x-x_i)^2+(y-y_i)^2}$ is the distance between position $(x,y)$
and the average position $(x_i,y_i)$ of the population density $f_i$.
The average positions have been chosen in the middle of the simulation area,  $x_i \!=\! y_i \!=\! 25$.
The function (\ref{fi2}) defines a population localized around $(x_i,y_i)$ within a radius $R_i$
with constant density $f_i \!=\! \mathcal{N}_i$,
smoothly decreasing to zero for $r_i(x,y) \!>\! R_i$ on a scale $\sigma_i$.
We have ensured that the density (\ref{fi2}) 
is nowhere larger than the carrying capacity, $f_i(x,y,t_0) \le K$.
In both examples C and D, population 2 starts from the same initial density (\ref{fi2})
with $\mathcal{N}_2 \!=\! 0.0163647$, $R_2 \!=\! 4$, and $\sigma_2 \!=\! 2$.
Instead, for population 1 two different initial conditions are used.
In example C it is given by the density (\ref{fi2}), 
with parameters $\mathcal{N}_1 \!=\! 0.0020048$, $R_1 \!=\! 10$, and $\sigma_1 \!=\! 3$;
In example D, the parameters are $\mathcal{N}_1 \!=\! 0.000548599$, $R_1 \!=\! 25$ and $\sigma_1 \!=\! 10$;
with such large values of $R_1$ and $\sigma_1$, the initial density $f_1$ in example D is practically uniform.
The parameters $\mathcal{N}_1$ and $\mathcal{N}_2$ have been chosen in order to have the same total initial populations
in both examples C and D, namely, $N_1(t_0) = 1.36$ and $N_2(t_0) = 2.64$.

From Fig.~\ref{multiplot-IC-growth1} one observes that,
differently from what happens in examples A and B (Fig.~\ref{multiplot-IC1}), 
increasing the initial spread of a population density favors its survival:
a more localized initial density $f_1$ (example C in Fig.~\ref{multiplot-IC-growth1}) is observed to disappear,
while a uniform initial condition for $f_1$ (example D in Fig.~\ref{multiplot-IC-growth1}) leads to its survival.
This can be traced back to the fact that in example D a higher growth of population 1 takes place in the peripheral regions, 
where the population density $f_2$ is negligible.
The time evolution of the total population $N_1(t)$ and $N_2(t)$ 
for examples C and D are depicted in Fig.~\ref{average-IC-growth1}.

\begin{figure}[ht]
 \centering
 \includegraphics[width=8cm]{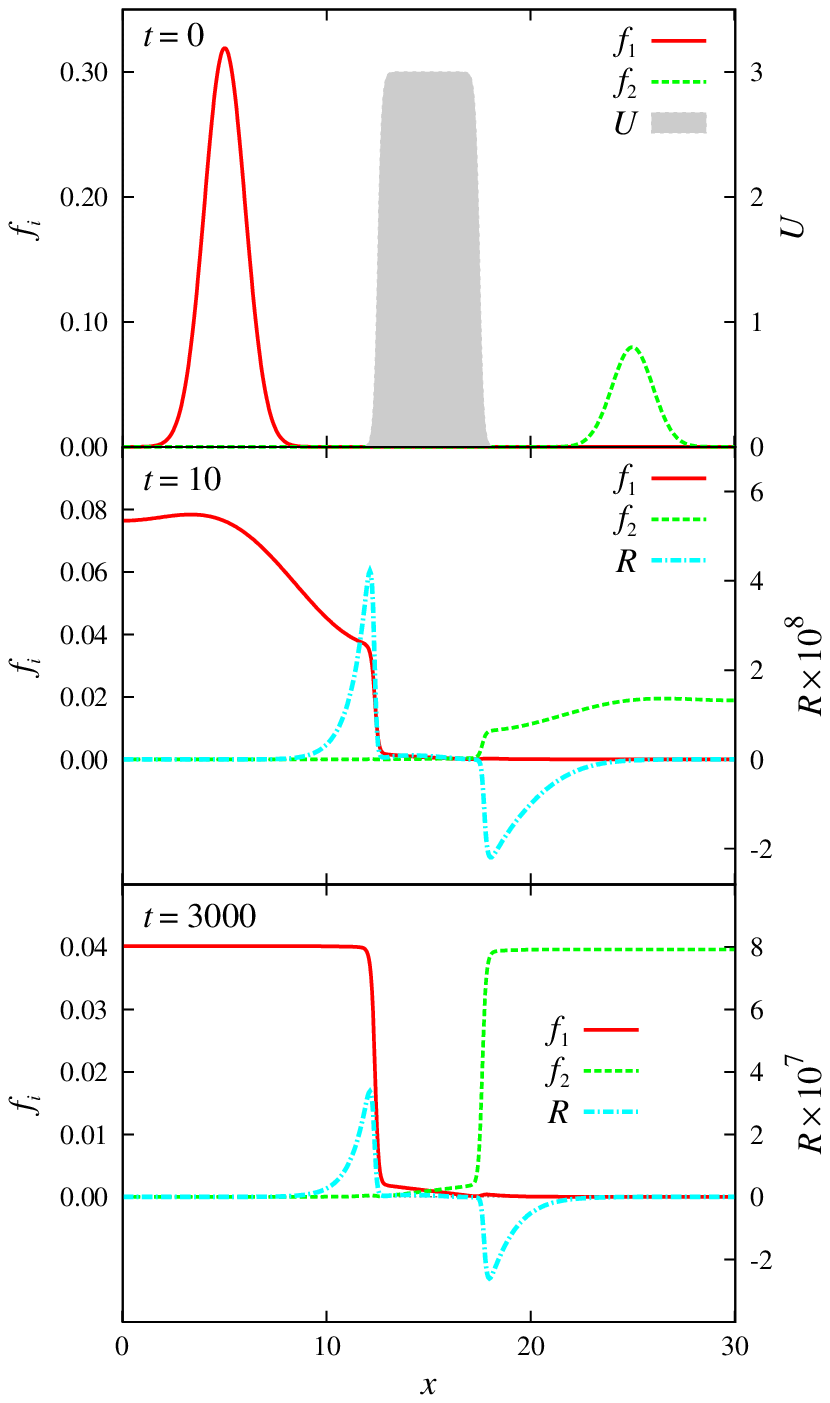}
 \caption{\label{barrier1-multiplot} 
 Evolution of the speaker population density $f_1(x,t)$ (continuous line, left axis) and $f_2(x,t)$ (dashed line, left axis)
 in the presence of a barrier $U(x)$, drawn at $t = 0$ (top, gray area, right axis).
 Language 1 is favored both in status ($s_1 \!=\! 0.6 \!=\! 1 \!-\! s_2$) and initial population 
 [$N_1(t_0) \!=\! 0.8 \!=\! 1 \!-\! N_2(t_0)$].
 At times $t \!=\! 10$ (middle) and $t \!=\! 3000$ (bottom)
 also the local reaction rate $R(f_1(x,t),f_2(x,t))$ is depicted (dashed-dotted line, right axis).
 In the asymptotic state (bottom) both languages survive, being localized on the opposite sides of the barrier.
 See text for further details.
}
\end{figure}

\section{Influence of boundary conditions}
\label{bc}
%
As shown for instance in the study of the three-state voter model in Ref.~\cite{Hadzibeganovic2008a},
boundary conditions can have a crucial influence on the competition process between cultural traits.
Here we investigate the problem of language competition comparing the influence of reflecting and periodic boundary conditions.
Using reflecting boundaries one can model a geographical area which is isolated, 
i.e., it is not possible to enter or leave it.
Periodic boundaries are a numerically convenient way 
to simulate an open region in the middle of a much larger accessible area.
We found that if the growth rate is negligible ($\alpha \approx 0$) 
the vicinity of reflecting boundary conditions favors the survival of a language.
For high growth rates, the effect of different boundaries is less appreciable.
Here we present the results for the case with no growth ($\alpha \!=\! 0$).
In Fig.~\ref{bc-multiplot} we compare the time evolutions of population densities $f_i(x,y,t)$
for two languages in the presence of reflecting and periodic boundaries.
For the language status, the values $s_1 \!=\! 0.55$ and  $s_2 \!=\! 0.45$ have been assigned.
In both examples the initial distributions $f_i(x,y,t_0)$ 
are assumed to have the Gaussian shape defined by Eqs.~(\ref{fi}), 
with $x_1 = 20$, $y_1 = 30$, $\sigma_1 = 3$, for population 1, and 
$x_2 = 45$, $y_2 = 5$, $\sigma_2 = 1$, for population 2;
the normalization constants are $N_1(t_0) = 0.37$ and $N_2(t_0) = 0.63$.
The size of the simulation area is $50 \times 50$ with $\Delta x \!=\! \Delta y = 1$,
the time step is $\Delta t \!=\! 0.01$, and the reaction constant $k = 1000$.

\begin{figure}[ht]
 \centering
 \includegraphics[width=8cm]{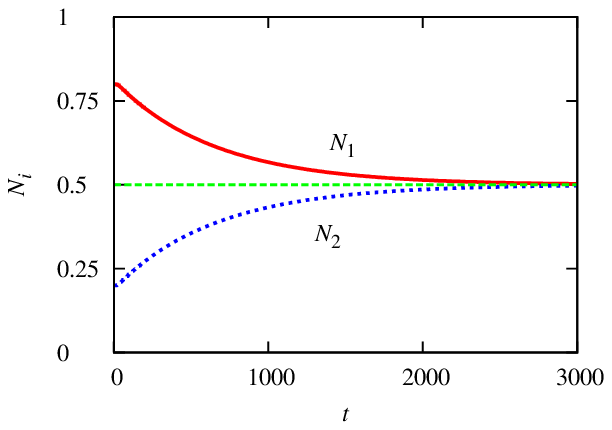}
 \caption{Time evolution of total populations $N_1(t)$ and $N_2(t)$ in the presence of a barrier,
 example of Fig.~\ref{barrier1-multiplot}.
 \label{barrier1-average}
}
\end{figure}

From Eq.~(\ref{cond}) one obtains,
for $s_1 \!=\! 1 \!-\! s_2 \!=\! 0.55$,
that the critical fraction ensuring survival of language 1 is 
$N_1^*\approx 0.36$ ($N_2^* \!=\!1\!-\!N_1^*\!\approx\! 0.64$).
Thus, the initial population fractions $N_1 \!=\! 0.37$ and $N_2 \!=\! 0.63$ used here would give 
a slight advantage to population 1 in the uniform case.
Instead, as one can see from Fig.~\ref{bc-multiplot}, 
language 1 disappears with reflecting boundary conditions (Fig.~\ref{bc-multiplot}, left), 
while it prevails if periodic boundary conditions are used (Fig.~\ref{bc-multiplot}, right).
This effect is due to the fact that reflecting boundaries bounce back 
a part of population 2  located near the boundary increasing the corresponding density $f_2$.
With open or periodic boundaries, 
population 2 would spread and its density $f_2$ decrease.
This in turn would lower the term representing the $2 \to 1$ 
language switching rate in the function $R(f_1,f_2)$.
Instead, a higher density $f_2$ favors the switching of speakers 1 to language 2. 
In Fig.~\ref{bc-average} the total populations $N_1(t) \!=\! 1 \!-\! N_2(t)$ 
for the two examples with reflecting and periodic boundary conditions are plotted.

\section{Geographical barrier}
\label{barrier}
%
In real situations the coexistence of more than one language 
in neighboring areas for long times is often observed~\cite{Nichols1997a}.
Some models, such as the AS model, describe situations in which
after a relatively short time only one of the competing languages survives.
Survival of different languages in the same area was shown in Ref.~\cite{Pinasco2006a}
to be possible when population growth is taken into account.
Another mechanism was suggested in Ref.~\cite{Mira2005a}, 
based on the similarity of the competing languages.
In the framework of the bit-string language evolution model,
the influence of a barrier on language diffusion and evolution was considered 
by Schulze and Stauffer, who showed that two different languages 
can exist on opposite sides of the barrier~\cite{Schulze2007a,Schulze2006c}.

In the present paper we concentrate on the role of purely geographical factors.
The scheme employed is different from that of Ref.~\cite{Patriarca2004c},
where the coexistence of two languages in neighboring regions was made possible 
by a barrier (geographical boundary or political border) 
affecting the form of the switching rate $R(f_1,f_2)$.
Instead, the mechanism described in the examples presented in Secs.~\ref{barrier} and \ref{island} below
is based on the presence of a geographical barrier which influences \emph{solely} population dispersal, 
while the cultural interaction remains the same as in the AS model.

\begin{figure}[ht]
 \centering
 \includegraphics[width=8cm]{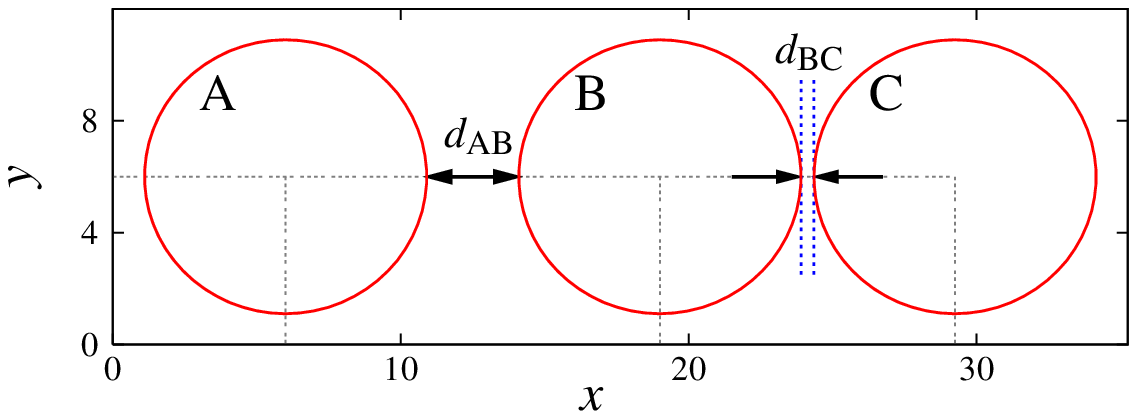}
 \includegraphics[width=10cm]{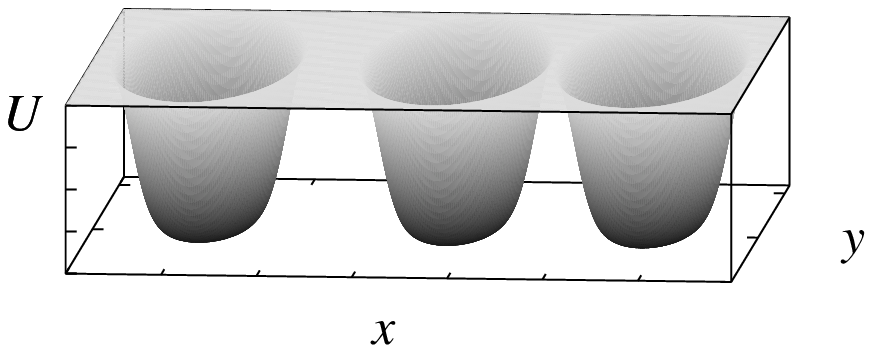}
 \caption{Schematic map (top) and potential energy landscape 
 representing the sea barrier (bottom)
 for the three-island configuration.
 Islands A, B, and C have a circular shape with the same radius,
 defined by the effective potential (\ref{U2}).
 Islands B and C are closer to each other than islands A and B.
 See text for further details.
 \label{fig_island}
}
\end{figure}
\begin{figure*}[ht]
 \centering
 \includegraphics[width=13.0cm]{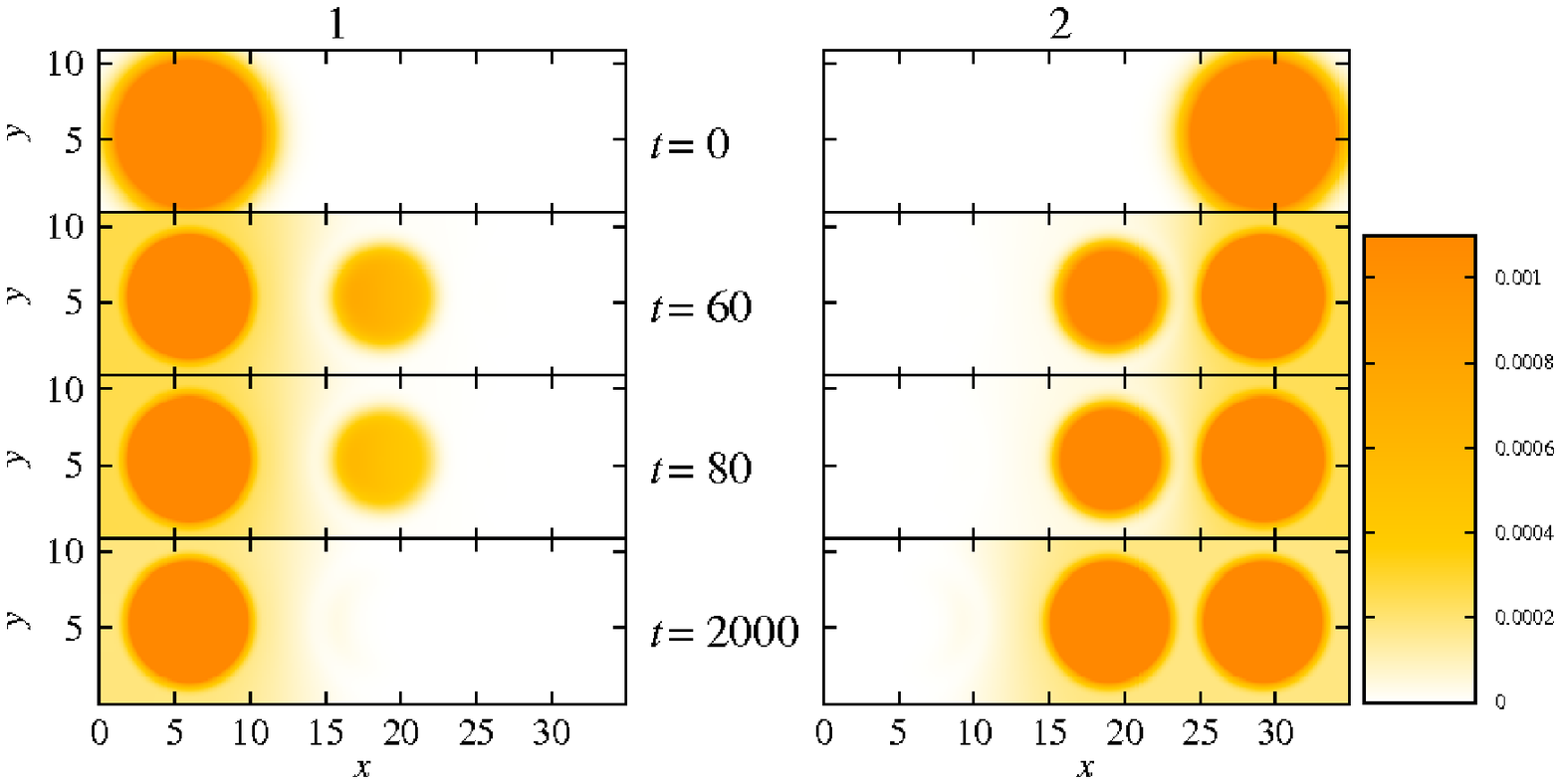}
 \caption{Evolution of population density $f_1(x,y,t)$ (left column) and $f_2(x,y,t)$ (right column).
  Notice the temporary presence of population 1 on (the central) island B at $t \!\approx\! 60$.
  Despite the lower status of language 2, geographical inhomogeneities favor 
  its immigration to the central island B.
  See text for the details.
 \label{fig_multiplot-island}
}
\end{figure*}

As a first example of geographical inhomogeneity, we consider a barrier,
representing e.g. a mountain chain, which divides the accessible area into two regions.
For the sake of simplicity we model the problem in one dimension.
We assume that the populations of speakers of language 1 and 2 are initially localized
on the opposite sides of the barrier, as depicted in Fig.~\ref{barrier1-multiplot} top.
We are interested in the influence of the barrier on the asymptotic state.
In order to answer this question, we have evolved the population densities $f_i(x,t)$ 
according to Eqs.~(\ref{genmodel1}), where $\nabla \!=\! \partial/\partial x$; 
we do not take into account population growth ($\alpha\!=\!0$).
Numerical integration is performed through the Crank-Nicolson method~\cite{Press1986a},
assuming reflecting boundary conditions.
A space step $\Delta x \!=\! 0.05$, a time step $\delta t \!=\! 0.001$, and a reaction constant $k \!=\! 200$, are used.
The barrier is modeled through the force field $\mathbf{F} \!=\! F(x) \!=\! - \partial U(x)/\partial x$,
where the potential $U(x)$, depicted in Fig.~\ref{barrier1-multiplot} top at $t \!=\! t_0 \!=\! 0$ (gray area), is
\begin{equation}\label{U1}
   U(x) = \frac{U_0}{1 + \exp[-(x-x_a)/\sigma_U\,] + \exp[(x-x_b)/\sigma_U\,]}  \, .
\end{equation}
This function represents a step of height $U_0$ located in the interval between $x_a$ and $x_b$ ($x_a < x_b$)
and going to zero outside of it on a scale length $\sigma_U$.
The parameters employed are: $U_0 \!=\! 3$, $x_a \!=\! 12.5$, $x_b \!=\! 17.5$, and $\sigma_U \!=\! 0.1$.
For the initial densities $f_1(x,t_0)$ (Fig.~\ref{barrier1-multiplot} top, left side of the barrier, continuous line) and
$f_2(x,t_0)$ (Fig.~\ref{barrier1-multiplot} top, right side of the barrier, dashed line) 
a Gaussian shape was assumed,
\begin{equation}
f_i(x,t_0) = \frac{N_{i}(t_0)}{\sqrt{2\pi}\sigma_{i}} \exp [-(x-x_i)^2/2\sigma_i^2] \, .
\end{equation}
We choose $\sigma_1 \!=\! \sigma_2 \!=\! 1$ and for the average coordinates 
$x_1 \!=\! 5$ and $x_2 \!=\! 25$, located symmetrically respect to the barrier
and the reflecting boundaries at $x \!=\! 0$ and $x \!=\!30$.
The status of language  1 is $s_1 \!=\! 1 - s_2 \!=\! 0.6 $ and 
the initial population fraction $N_1(t_0) \!=\! 1 - N_2(t_0) \!=\! 0.8$.
Thus, language 1 is highly favored, 
regarding both status and initial total population.
In fact, if there is no potential barrier, language 2 is observed to disappear, as it is easy to guess.
In the presence of the potential barrier (\ref{U1}) the evolution of the population densities
is shown in Fig.~\ref{barrier1-multiplot}:
both languages survive in the asymptotic limit on the opposite sides of the barrier.
The time evolution of the total populations $N_i(t)$ is shown in Fig.~\ref{barrier1-average}.
The survival of language 2 on the right side of the barrier can be explained as follows:
the potential barrier modulates diffusion toward both the directions and in particular
it decreases the flux of population 1 from the left toward the right region.
This in turn causes the term $k \, s_1 f_1^{\, a}(x,t) f_2(x,t)$ in the reaction rate (\ref{R})
to remain very small in the right region,
so that the local $2 \!\to\! 1$ language switching rate is negligible respect
to that of the complementary process $1 \!\to\! 2$.

The survival of both languages observed in the example discussed above 
can be ascribed to the interplay between historical conditions 
(initial localization of the two speakers communities on opposite sides of the barrier)
and geographical constraints (presence of the barrier).
In Fig.~\ref{barrier1-multiplot} center and bottom also the switching rate $R(f_1,f_2)$ is depicted with a dash-dotted line.
The largest values of $|R|$ are located close to the barrier borders,
where speakers coming from the other side meet the local speakers and switch to the local language.
In Fig.~\ref{barrier1-average} one can also notice that the asymptotic total population values are equal
$N_1(t \!\to\! \infty) = N_2(t \!\to\! \infty) = 1/2$,
as a consequence of the identical dispersal properties assumed for the two populations
and the symmetrical geometry of the system.

\begin{figure}[ht]
 \centering
 \includegraphics[width=8cm]{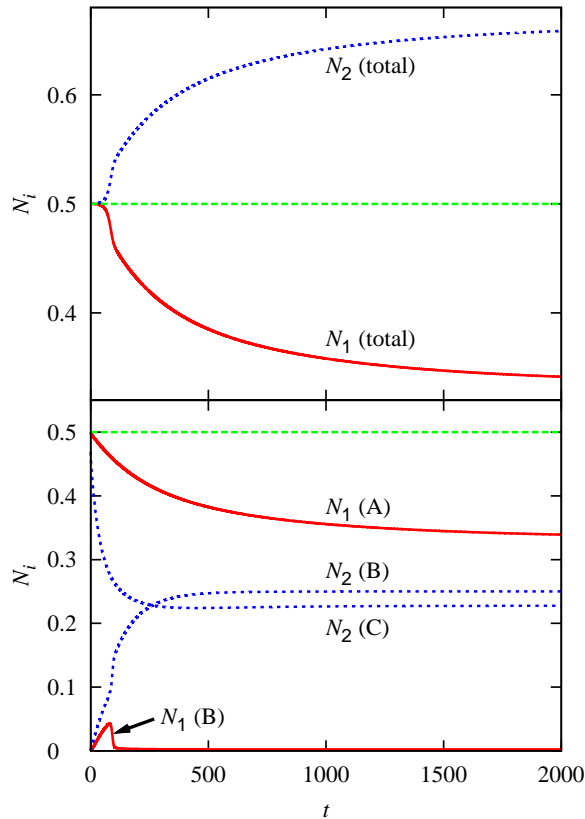}
 \caption{Time evolution of the speaker community 1 (continuous line) and 2 (dotted line)
 for the three island example.
 Top: total populations.
 Bottom: partial population 1 on island A and B and partial population 2 on island B and C (the partial populations not shown are negligible).
 Notice that both populations 1 and 2 are present on the central island B until $t \approx 100$.
 \label{fig_average-island}
}
\end{figure}

\section{Immigration to an island}
\label{island}
%
Let us now see, what happens if we consider the spreading of two languages toward the same initially empty region.
To make an example, we study two islands A and C, initially colonized 
by populations speaking language 1 and 2, respectively.
Language 1 has a higher status $s_1 \!=\! 1 \!-\! s_2 \!=\! 0.6$. 
Between the two islands A and C is located a third island B, which is empty.
For simplicity, we choose for the islands a circular shape with identical radius $R$;
the centers of the islands are located on a line.
The situation is illustrated in Fig.~\ref{fig_island} top.
The system has been studied on a rectangular simulation area of sides $L_x = 35$ and $L_y = 11$.
The population densities were evolved through the explicit Euler algorithm (\ref{genmodel2})
on a $350 \times 110$ lattice with steps $\delta x = \delta y = 0.1$,
using a time step $\delta t = 0.001$ and a rate constant $k = 2000$.

Speakers can freely diffuse inside the islands,
but have to overcome a barrier in order to cross the sea and reach other islands.
The effective potential $U(x,y)$ modeling the barrier due to the sea also defines the island shapes;
it is depicted in the lower part of Fig.~\ref{fig_island} and is given by
\begin{equation}\label{U2}
U(x,y) \!=\!
\left\{
  \begin{array}{ll}
    U_0 \exp\left\{-[\,r_j(x,y) - R\,]^2 / \, 2 \, \sigma_U^2 \,\right\},&r_j < R ,\\
    U_0 \, ,                                                             &\hbox{otherwise.}
  \end{array}
\right.
\end{equation}
Here $j = A, B, C$ labels the islands and $r_j = r_j(x,y) \! = \! \sqrt{(x-x_j)^2+(y-y_j)^2}$
is the distance between the generic position $(x,y)$ and the center $(x_j,y_j)$ of the $j$-th island;
i.e., a value $r_j(x,y) \!<\! R$ for a given $j$ corresponds to a point $(x,y)$ inside the $j$-th island,
while if $r_j(x,y) > R$ for all $j \!=\! \mathrm{A,B,C}$ the point $(x,y)$ is on the sea. 
The location of the three islands can be recognized in Fig.~\ref{fig_island} bottom as the three potential wells,
while the sea is represented by the plateau $U(x,y) \!=\! U_0$.
The parameter values used are: $\sigma_U \!=\! 2$, defining the smoothness of the potential step,
$U_0 \!=\! 4$, for the barrier height, 
and $R \!=\! 5$, for the island radius;
the coordinates of the centers of the island are
$x_\mathrm{A} \!=\! 6$, $x_\mathrm{B} \!=\! 19$, $x_\mathrm{C} \!=\! 29.25$,
and $y_\mathrm{A} \!=\! y_\mathrm{B} \!=\! y_\mathrm{C} \!=\! 6$.
The initial total populations on islands A and C have been assigned the same value $N_1(t_0) \!=\! N_2(t_0) \!=\! 1/2$.
The population densities $f_i(x,y,t_0)$ have the same Gaussian shape (\ref{fi})
with $\sigma_1 \!=\! \sigma_2 \!=\! 2$ and average coordinates $x_i$ and $y_i$ coinciding 
with the center coordinates of the respective islands.

Language 1, with the higher status $s_1 \!=\! 0.55$, would be favored in the absence of barriers.
We would like to know if on island B eventually prevails language 1 or 2.
It is clear that if the central island B is located symmetrically between islands A and C
then it is language 1 that in the end will prevail (also on island A) due to its higher status.
Population 2 may still survive on island C, where it was initially dominating,
if the sea represents a large enough barrier,
due to the effect described in the previous section which would transform islands C into a refugium.
If not, language 1 may prevail finally also on island C.

On the other hand, if it is easier to cross the B-C rather than the A-B channel, 
e.g. if island B is closer to island C than to A,
then language 2 may spread first on island B and then maintain its superiority thanks to the barrier
due to the presence of the A-B channel.
We have studied the problem for various values of the distance $d_\mathrm{BC}$ between island B and C,
while keeping constant the other distance $d_\mathrm{AB}$.
The example illustrated in Figs.~\ref{fig_multiplot-island} and \ref{fig_average-island}
corresponds to a value $d_\mathrm{AB} \!=\! 3$ and to a much smaller distance $d_\mathrm{BC} \!=\! 0.25$ between islands B and C.
Figure~\ref{fig_multiplot-island} shows how populations 1 and 2 disperse 
over the neighboring island starting from their initial locations.
Due to the geographical asymmetry favoring dispersal of population 2,
there is a much larger flow of population 2 from island B to C than population 1 from island A to B.
This in turn leads to a rapid spreading of language 2 on island B.
Once language 2, with a lower status, dominates on island B, 
the wide sea barrier between island A and B will maintain the advantage gained by language 2.
Figures~\ref{fig_multiplot-island} and \ref{fig_average-island} also show a small presence of language 1 
on the central island B limited to a short time interval.

\section{Conclusion}
\label{conclusion}
%
Dispersal in space and time of two languages or cultural traits 
competing in the same geographical area has been studied 
through an extended language competition model based on the one proposed by Abrams and Strogatz.
We have discussed how initial and boundary conditions, as well as geographical inhomogeneities,
have a relevant (even drastic) meaning for language spreading and competition.

We have observed various examples where a language, 
which in the corresponding homogeneous model would disappear, actually survives.
Namely, in a homogeneous model (without space dimensions),
for given values of the parameters,
the evolution of a population is determined by the total initial population.
In the diffusion model studied here, the same total population of speakers, 
initially distributed in space in different ways, 
may evolve toward opposite asymptotic scenarios.
Another result of our investigation shows that, when growth is negligible, 
a language, whose dispersal is more affected by the geographical boundaries, 
is favored respect to the case when no limiting boundaries are present.
In the presence of geographical inhomogeneities, modeled as potential energy barriers,
languages can survive in different regions, 
despite the possibly lower status and smaller initial populations.
The effects discussed in the present paper are purely geographical, 
in the sense that they are related to the diffusion processes
and to the modulation of diffusion due to inhomogeneities of the background.
They influence culture spreading only indirectly and are not due to a change in the cultural interaction law, 
differently from the model introduced in Ref.~\cite{Patriarca2004c}. 

The diffusion model and the highly idealized examples presented in this paper 
are intended as a first step toward a quantitative description 
of the space-time diffusion of language and cultural traits.
Our study will hopefully be useful in solving some of the many challenging problems 
regarding language diversity~\cite{Nichols1997a}.
At the same time, a more detailed understanding of the mechanisms underlying culture transmission 
could be valuable concerning the alarming rate of disappearance of cultural and linguistic diversity.

\begin{acknowledgments}
This work was supported by the Estonian Science Foundation Grant No. 7466.
M.P. gratefully acknowledges also financial support from the GIACS Workshop on {\it Language Competition} 
(2006 September 11-14, Warsaw, Poland)
and the participants of the workshop for useful discussions and constructive remarks.
\end{acknowledgments}


\end{document}